\documentclass[final,3p,twocolumn,times]{elsarticle}

\usepackage{amssymb}
\usepackage{graphicx}
 \usepackage{threeparttable}
 \usepackage{multirow}
\usepackage[usenames]{color}
\usepackage{cancel}
\usepackage[normalem]{ulem}
\usepackage{amsmath}
 \usepackage[dvipdfm, pdfstartview=FitH]{hyperref}
\hypersetup{colorlinks,
        linkcolor=black,
        filecolor=black,
        urlcolor=blue,
        citecolor=black,
        pdftitle={For Alstom},
        pdfauthor={             },
        pdfsubject={            },
        pdfkeywords={           },
        pdfproducer={ps2pdf} }

\begin{document}

\begin{frontmatter}

\title{
Cyclic motions in Dekel-Scotchmer Game Experiments
}

\author{
Zhijian Wang
}

\cortext[cauthor]{email: wangzj@zju.edu.cn }
\address{
 Experimental Social Science Laboratory, Zhejiang University,
Hangzhou, 310058, China
}
\address{
State Key Laboratory of Theoretical Physics, Institute of Theoretical Physics, Chinese Academy of Sciences, Beijing, 100190, China
}
\date{\today}

\begin{abstract}
TASP (Time Average Shapley Polygon, Bena{\=\i}m, Hofbauer  and Hopkins, \emph{Journal of Economic Theory}, 2009), as a novel evolutionary dynamics model, predicts that
a game could converge to cycles instead of fix points
(Nash equilibria).  To verify TASP theory, using the four strategy Dekel-Scotchmer games (Dekel and Scotchmer,  \emph{Journal of
  Economic Theory}, 1992),
%four Rock-Paper-Scissors-Dumb
four experiments were conducted (Cason, Friedman and Hopkins,  \emph{Journal of
  Economic Theory}, 2010), in which,
% and gave some positive results, h
 however,
% the previous work
 reported no evidence of cycles (Cason, Friedman and Hopkins, \emph{The Review of Economic Studies}, 2013).
% which is a puzzle.
We reanalysis the four experiment data by testing the stochastic averaging
 of angular momentum in period-by-period
  transitions of the social state. We find, the existence of persistent cycles in Dekel-Scotchmer game can be confirmed.
  On the cycles, the predictions from evolutionary models had been supported by the four experiments. \\
%  , together with
% extensively observed cycles in different game experiments,
% suggested that,
% % evolutionary models could apply not only to animals with genetically heritable strategy, but also to social strategy evolution.
%the expected motions in evolutionary dynamics can be verified by the actual motions in laboratory experiments quantitatively.\\
  \end{abstract}

\end{frontmatter}

\tableofcontents
%\newpage
\section{Introduction}

\begin{figure}
  \begin{center}
          \includegraphics[width=1\linewidth]{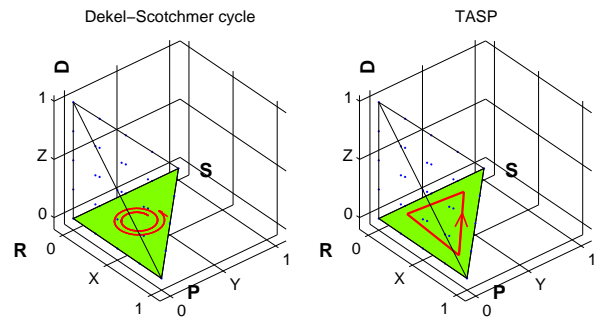}
          %C:\Users\ibm\Documents\MATLAB\TASP2010\illustrateDekelState3DE_Sapce.m
  \end{center}
  \caption{Cyclic trajectory and its angular momentum. (a) Ideal Dekel-Scotchmer cycle (refer to the Figure 1 in p396 in  \cite{dekel1992evolution})
   and (b) Time Average Shapley Polygon (TASP) (refer to the Figure 1 in p2313 in \cite{friedman2010tasp})
  for unstable RPSD game.
  The frequencies of strategies $P$ and $S$ are on the horizontal axes and of strategy $D$ on the vertical axis. The red arrows indicate the accumulated angular momentum (vector) respects to the cycles.
%  using the measurement defined in Eq.~\ref{eq:AM}.
   \label{fig:DekelandTASP}
  }
\end{figure}

While facing a game, the first step is to look for Nash equilibra (fixed points)~\cite{nash1950equilibrium}, but in some condition, instead of fix points, a game could converge to  cycles. As an example in evolutionary game theory catalog,
recently, a dynamic theory --- Time Average Shapley Polygon (TASP) theory~\cite{benaim2009learning} ---
% {\color{red} published in a paper in this journal},
is built
%and gives
giving
a precise prediction about non-equilibrium play in games.
%It predicts that: I
%So, the cycle is expected.
To test TASP theory, four exemplified experiments of the Dekel-Scotchmer game \cite{dekel1992evolution}, called also as Rock-Paper-Scissors-Dumb (RPSD) games \cite{friedman2010tasp},
were conducted  by Cason \emph{et al}.~\cite{friedman2010tasp}.
The four  experiments game were multi round repeated, set as discrete time (instead of continuous time  as \cite{Friedman2012}), at the same time the matching protocol are randomly pair-wised  which called as evolutionary protocol \cite{Huyck1999,Samuelson2002}. In the four experiments,  some evidences supporting TASP are found. But no cycle is reported in \cite{friedman2010tasp}, which is emphasized by the authors in their recent literature \cite{Friedman2012}. In fact, this is the second time for the cycles in the Dekel-Scotchmer game was declined in experiments. This game had firstly been tested in experiments, no cycles was found which was also emphasised by the authors \cite{Huyck1999} (We go back this point in  discussion, section \ref{sec:Discuss}). All these seems to suggest that there is no cycle in the four Dekel-Scotchmer game experiment.

As illustrated in Fig. \ref{fig:DekelandTASP}, cycle should along $R$, $P$, $S$, $R$, $P$ ... in the $RPS$ plane in the tetrahedronthe state space of the Dekel and Scotchmer game. Since the game was designed \cite{dekel1992evolution}, not only TASP,
%not only the designers of this game \cite{dekel1992evolution} and TASP theory~\cite{},
variants evolutionary models  has expected the cycle \cite{benaim2009learning,dekel1992evolution,Weibull1997,Sandholm2011,hauert2002volunteering}.
%
%in these games has predicted the cycle
%along RPS plane in long run.
%This is a puzzle, because:
%
%\textbf{(b)}
%Empirically, biologies exhibit the existence of evolutionary cycles \cite{Sin96,Kir04}.
Empirically, as in biology system \cite{Sin96,Kir04}, in experimental economics,
%Difference from common wisdom that
%evolutionary models could only apply to animals with genetically
%heritable strategy, e
evolutionary models have been supported extensively~\cite{Plott2008,Samuelson2002,Friedman1996,Huyck2008,Nowak2012,berninghaus2003power}.
Recently, in the discrete time  protocol,
the cycles have been constantly tested out \cite{XZW20130621,XuWang2011ICCS, XuWang2010XM,
wang2012evolutionary,Xu2013,XuWangAsymmetryRPS,
XWMm2011,WangXu2011,xu2012test,xu2012periodic}.  % by Xu \emph{et. al.} in their own standard matching-pennies   and standard RPS game experiments, and also in others' experimental data in existing literatures \cite{Binmore2001,selten2008}.
%In \cite{friedman2010tasp}, using same protocol, why the social does not cycle?
% However, as point out by a recent paper reporting cycles in continuous time RPS experiments~\cite{Friedman2012},  their paper on the four game experiments~\cite{friedman2010tasp} \emph{reports no evidence of cycles}.
So, it is enigma that  using the similar protocol in the four experiments, why does the RPS cycle not exist.  The objective of this paper was to study whether or not cycle exists the four Dekel-Scotchmer game experiments.

%To solve this puzzle, angular momentum ---
%an observation of rotation in classical physics ---
%is employed\footnote{This measurement has been used to index
%cycle in 2 $\times$ 2 games in the seven experiments from
%Binmore \emph{et al}. (2001)\cite{Binmore2001} and using the stochastic averaging $L$ to
%compare with models \cite{wang2012evolutionary}.} to measure the cycles
%in the period-by-period transitions (PPT) of social state
%in the experiments~\cite{friedman2010tasp}.

%On the cycles¡¯ existence, direction, strength and persistence, it is
%noticed that, the predictions from TASP theory had been supported by the four experiments.
%%These observed cycles,
%These results suggest that, the expected motions in
%evolutionary dynamics can be verified by the actual motions in laboratory experiments quantitatively.

%\section{A brief review on the RPSD experiment}\label{sec:introTASP}

\begin{figure}
  \begin{center}
   \includegraphics[width=0.8\linewidth]{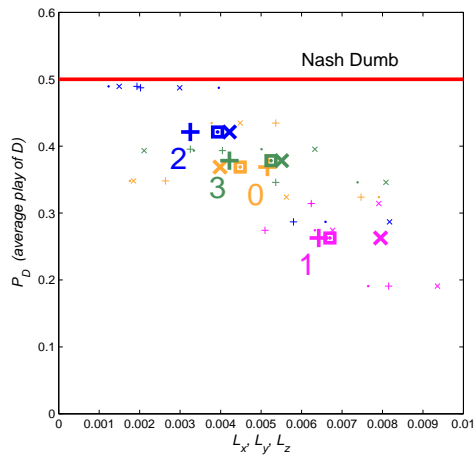} %plotLxLyLz_DfromEXCEL
  \end{center}
  \caption{
     The relation between observed average angular momentum ($\bar{L}_x, \bar{L}_y, \bar{L}_z$) and observed average play of $D$ ($P_D$, see also the Table 1 in \cite{friedman2010tasp}). Observed ($\bar{L}_x, \bar{L}_y, \bar{L}_z$) can be obtained from $k$=1 rows in Table~\ref{tab:k1-k4p}. Red line is Nash equilibrium Dumb ($D$-play). Color [yellow, purple, blue, green]  presents game-[0,1,2,3], respectively. Symbol $(\square,\times,+)$ represents $(\bar{L}_x,\bar{L}_y,\bar{L}_z)$  respectively. Enlarged  symbols are the averages by games and smaller by sessions.
  \label{Dumb_LxLyLzByGame}
          }
\end{figure}

 The four experiments have a 2 $\times$ 2 design
 \cite{friedman2010tasp} shown in Table \ref{tab:4game}. The first design is the two game matrix. Both settings are Dekel-Scotchmer game constructed from the Rock-Paper-Scissors (RPS) game with the addition of a fourth strategy called Dumb ($D$). The payoff matrix can be presented as
\begin{center}
 $ \begin{array}{lrrrr}
  & ~~~~R& P & S & D \\
\end{array}$\\
 $ \begin{array}{l}
   R \\ P \\ S \\ D \\
\end{array}$
$\left[\begin{array}{rrrr}
  a & 0 & b & c \\
  b & a & 0 & c\\
  0 & b & a& c\\
   d & d & d& 0
\end{array}
\right]$.
\end{center}
  For Unstable and Stable games, $[a~b~c~d]$ equals to [90 120 20 90] and [60 150 20 90], respectively.
 %$ \left[
%\begin{array}{rrrr}
% 90 & 0 & 120 & 20 \\
% 120 & 90 & 0 & 20\\
% 0 & 120 & 90& 20\\
% 90 & 90 & 90& 0
%\end{array}
%\right]$
%\\ and for stable game is \\ RPSD$_S$ = $\left[
%\begin{array}{rrrr}
% 60 & 0 & 150 & 20 \\
% 150 & 60 & 0 & 20\\
% 0 & 150 & 60& 20\\
% 90 & 90 & 90& 0
%\end{array}
%\right]$. \\
 Both games have the same
unique Nash-Dumb (the probability to choose Dumb is 1/2,
shown as the redline in Fig.~\ref{Dumb_LxLyLzByGame}).
The second design is two conversion rates of Experimental Francs (the entries in the game matrix) to US Dollars. In the High-pay (Low-pay) treatment, 100 EF = \$5 (\$2). In High-pay games, the monetary incentive for optimal strategy is stronger and less noise is expected. Mainly,
%\footnote{There have some more consideration (like belief testing) in the experiment for comparing with more models (like QRE and EWA).},
the settings of the four game are summarized as shown in Table \ref{tab:4game}. While these games are \emph{identical} in their equilibrium predictions, they differ quite substantially
in terms of predicted learning behavior
%.
%Theoretically, if all use a fictitious play-like learning process to update their play \cite{benaim2009learning},
The stable games (game-2,3) would converge  to the Nash
equilibrium. At the same time, in the unstable  game (game-0, 1), play will approach to a cycle (in RPS-plane, see green triangle in Fig. \ref{fig:k1-k4}) in which there would be no weight placed on the strategy Dumb ($D$).
So,  the correction of TASP can be evaluated \cite{friedman2010tasp} by the \emph{average play} of $D$ ($P_D$). The main result \cite{friedman2010tasp},
as illustrated in $y$-axis in Fig~\ref{Dumb_LxLyLzByGame},
$P_D$ in game-1 leaves Nash Dumb the farthest and game-2 the closest.
These meet TASP theory well.

Cycles are also expected by TASP theory (Fig.\ref{fig:DekelandTASP}) in these four games.
%which are correlated to the deviation from Nash equilibrium.  For the four experiemtns,  t
The theoretical arguments of the fictitious learning model have been well analyzed \cite{friedman2010tasp} ------  For the game treatment, as the \emph{basic argument} (p.2312 in~\cite{friedman2010tasp}),
% if all use a fictitious play-like learning process to update their play,
 in the stable game there would be convergence
 to the Nash equilibrium. In the unstable game, however, there will be divergence from equilibrium and play will approach a
cycle in which no weight is placed on the strategy Dumb (D). For the payoff treatment, high-pay has the  same  effect as an \emph{increase} in the noise parameter (p.2317 in~\cite{friedman2010tasp}) ------
 Accordingly, the quantitative expectation is that the more deviation from equilibrium, the more cyclic motion will be, which forms the second testable point following presented by Table~\ref{tab:4game}. For the objective of this paper, the expectation on cycle
%is called --- RPS cycle hypothesis (RPS-CH).  The hypothesis
 could be decomposed into three testable hypothesis:
\begin{flushleft}
 (1) Cycles exist only along $R,P,S,R$,... direction
 shown in Fig.~\ref{fig:k1-k4} and Table~\ref{tab:k1-k4}. Explanation for this testable points see section \ref{sec:testTables}. The results, see section \ref{sec:cycleDir}, support this point.\\
 (2) Cycles' strength depends on games, shown  in Table~\ref{tab:4game},
  in which game-1 is the largest.  Results, see section \ref{sec:cycleStrength}, support this point. \\
 (3) Cycles' persistence depends on games,
  in which game-1 performs the best.  Results, see section \ref{sec:CyclePerisistenceTASP}, support this point.
\end{flushleft}

 These three hypothesis are tested in this paper. Using angular momentum (an observation of rotation in classical physics),
%is employed
%\footnote{This measurement has been used to index
%cycle in 2 $\times$ 2 games in the seven experiments from
%Binmore \emph{et al}. (2001)\cite{Binmore2001} and using the stochastic averaging $L$ to
%compare with models \cite{wang2012evolutionary}.}
in the period-by-period transitions (PPT) of social state
in the four experiments~\cite{friedman2010tasp}, we test the cycles
and find that all the three theoretical arguments are supported in significant.
% the
%existence of persistent cycles can be confirmed.
%As Dekel-Scotchmer cycle is widely concerned (e.g., \cite{benaim2009learning,Weibull1997,Sandholm2011}),
% but its existence has not been confirmed \cite{Friedman2012,Nowak2012,Xu2013,Huyck1999}. We
we hope our observation can provides an exemplified evidence of the existence of Dekel-Scotchmer cycles.
%The methods and the results are as follows.

\begin{center}
\begin{table}
\caption{\label{tab:4game}The four games (treatments)}
\begin{tabular}{|c|ccc|}
\hline
game i.d. &	~~~~Low pay~~~~ &	&~~~~High pay~~~~ \\
  \hline
~~~~Unstable~~~~&	\textcolor[rgb]{1.00,0.50,0.00}{\textbf{0}}& $ \leftarrow  $ &	\textcolor[rgb]{1.00,0.00,1.00}{\textbf{1}} \\
&	$\downarrow $  &  &	$\downarrow  $ \\
Stable&	\textcolor[rgb]{0.00,0.00,1.00}{\textbf{2}}& $ \leftarrow  $  &	\textcolor[rgb]{0.25,0.50,0.50}{\textbf{3}} \\
  \hline
\end{tabular} \\
\scriptsize
Each game has 3 repeated sessions with 12 subjects in each session.
Game in each session is about 80 times repeated.
Matching protocol is randomly pairwise.
%Unstable/Stable is determined by game matrix and Low-Pay/High-Pay is determined by exchange rate.
On cycle expected by RPS-CH, $a \rightarrow b$ presents the  strength in game-$a$ larger than that in game-$b$ ($\bar{L}_a > \bar{L}_b$). The related  empirical results are shown in Table \ref{tab:LxyzZval}.
\end{table}
\end{center}
\section{Methods}\label{sec:Lmeasurement}
 \subsection{State space settings}

\begin{figure}
  \begin{center}
          \includegraphics[width=1\linewidth]{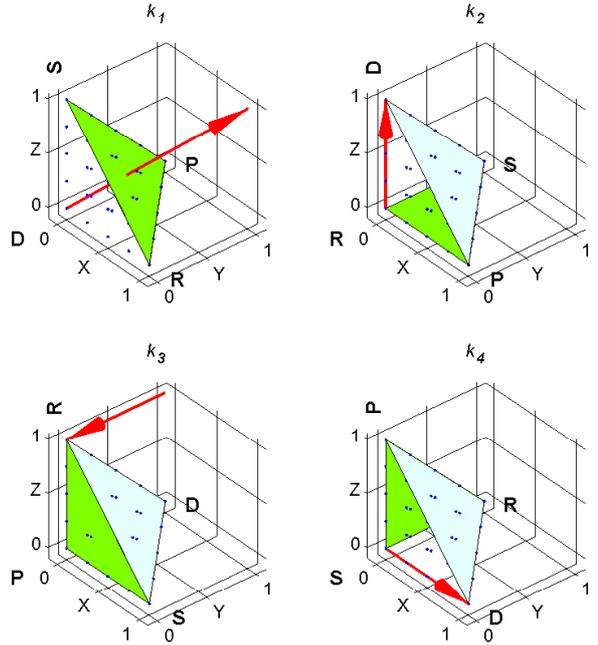}
  \end{center}
  \caption{
    Social state space of the RPSD game in 3D representation.
      Each social state is represented by a (blue) dot. Each $k_1,k_2,k_3,k_4$ setting, see the $x-y-z;O$-column in  Table~\ref{tab:k1-k4},  is illustrate as a sub-figure. The green triangle is the RPS-plane. The red arrow indicating $L$-direction if the net notations are along $R\!\rightarrow\!P\!\rightarrow\!S\!\rightarrow\!R$ (expected by TASP theory see the ($L_x,L_y,L_z$)-columns in Table~\ref{tab:k1-k4}). For example, in $k_2$ setting, the red arrow is (0,0,1) which means ($L_x=0, L_y=0, L_z>0$) are expected.
   \label{fig:k1-k4}
  }
\end{figure}

There are four pure strategies in the game, therefore
we use a four dimensional (4D) vector $(x, y, z, u)$ to denote a generic \emph{social state} of
the population, where $x$, $y$, $z$ and $u$ are
the fractions of players using strategy $R$, $P$ $S$ and $D$, respectively \cite{XuWang2011ICCS,Nowak2012}.
The fraction must be one element in ($0,1/N,2/N...,1$) set in an $N$-players game.
There are 4 pure social states which can be denoted as
$W_i$ ($i\!\in\!\{R,P,S,D\}$).
% and  $R_0$=(1,0,0,0), $P_0$=(0,1,0,0), $S_0$=(0,0,0,1) and $D_0$=(0,0,0,1)).
At the same time,
the sum of the fractions is 1 ($x+y+z+u$=1),
the 4D space is constrained. Hence, it can be projected into a 3D space.

By the % {\color{red} permutating ? permutating or permuting}
permuting
 $D, R, P, S$  at $O$=(0,0,0) and other three at $x,y,z$-axis respectively, there could have four ways (denoted as $k_1,k_2,k_3,k_4$) to realize the projections.
%The assignments see column-($x,y,z;O$) in Table~\ref{tab:k1-k4}.
See column-($x,y,z;O$) in Table~\ref{tab:k1-k4} for the assignments.
The four 3D spaces can be presented graphically as a trirectangular tetrahedron lattice\footnote{
In the studied case of $N$=12 and each subject can choose one in four pure strategy in one period,
the total number of different observable social states
 is $ \prod_{i=1}^3 \left[i^{-1}(N+i)\right]$
 =
 455. These states form the state space (lattice).
}
as illustrated in Fig.~\ref{fig:k1-k4}.

 \subsection{Period-by-period transition (PPT)}

In such lattice space, generically speaking,  the observed social state
 depends on time. From one period ($t$) to its next period ($t+1$), one  social state transition from $x(t)$ to $x(t+1)$ , called as one period-by-period transition (PPT), can be observed. Each PPT is a 3D vector in the lattice space.
Successive PPT vectors form  an evolutionary trajectory in 3D. For example, in a 80 rounds experimental sessions, an evolutionary trajectory of 80 nodes in which there are 79 PPT can be obtained.

\subsection{Angular momentum in PPT}
%\href{http://www.pha.jhu.edu/~broholm/l20/node5.html}{Generalization of Angular momentum} \\
For simplicity we consider first a particle (with mass $m$=1) moving
with respect to
a specific reference point (denoted as $o$).
Consider one PPT, from $ x(t)$ to $x(t+1)$,
the instantaneous angular momentum vector $L(t)$ can be expressed as~\cite{wang2012evolutionary}
\begin{eqnarray}\label{eq:AM}
    L(t) &=&  \left[x(t) - o\right] \times  \left[x(t\!+\!1) - o\right], %\\
    %\omega &=& \frac{1}{T} \sum_{t=0}^{T-1}\!  \mathbf{Sgn}(L) \arccos \frac{x_t \cdot  x_{t+1}}{|x_t||x_{t+1}|},
\end{eqnarray}
%$W_{\tau}(t)=(1/{\tau})\int_t^{t+\tau}\!\left[V(t')/\left\langle V\right\rangle\right]dt'$
in which
%$x(t)$ is the 3D strategy vector at time $t$,
%$o$ is a given reference point and
 symbol $\times$
means cross product of the two vectors. So, $L(t)$ has a magnitude equal to the area of the parallelogram with edges $\left[x(t) - o\right]$ and $\left[x(t\!+\!1) - o\right]$, has the attitude of the plane spanned by $\left[x(t) - o\right]$ and $\left[x(t\!+\!1) - o\right]$, and has orientation being the sense of the rotation that would align $\left[x(t) - o\right]$ with $\left[x(t\!+\!1) - o\right]$. It does not have a definite location or position \cite{hestenes1999new}. No lose of the generality, the random mixed (1,1,1,1)/4 is used as the reference point to report the following results.\footnote{$L$ vector is reference point ($o$) depended in one PPT. It is no difficult to prove that, $\bar{L}$ of a closed loop is independent of reference point setting. To test the robustness of the results in Table~\ref{tab:k1-k4p}, Table \ref{tab:k1L39L41} and Table \ref{tab:LxyzZval}, the reference point has been set for all the 455 states, respectively. The results are consistent.}

This measurement can be interpreted as following. In equilibrium, in long run, the time average of $L$ (denoted as $\bar{L}$) is 0,
%$\displaystyle\lim_{t\rightarrow \infty} \bar{L}=0$
because of the detailed balance in PPT \cite{XZW20130621,young2008stochastic}.
In non-equilibrium,  $\bar{L}$ should be significantly different from zero which provides
combinative cycles' information (strength and
direction of rotation) of the "\emph{tumbling cycles}".
This way is to proceed from the
microscopic level motions to
the macro level observation by \emph{stochastic averaging}
\cite{deng2004stationary,zhu2005stationary,
deng2009some,cai2007stochastic} over PPT.

%Hence, to test cycles turns to test the stochastic averaging $\bar{L}$.
%\footnote{For the comprehensibility of this measurement, an animation is shown in web page \href{http://socexp.zju.edu.cn/}{http://socexp.zju.edu.cn/Anim4AngMom}}.
%\subsection{Components of $L$ as test points}
% Using $L$, one would be able to trace cyclic motion in PPT
%with any given $o$.

In our studies case, there are three components ($L_x, L_y,L_z$) of an $L$.
Each component describes the rotation along its own
directions respectively. So, each PPT can provide
one sample in each of the three directions respectively.
%of observation on $L$\footnote{
In sufficient samples,
if a component $\bar{L}_w$ ($w \in \{x,y,z\}$) deviates from 0 with the statistic
significance, cycles exist in the direction.

As mentioned above, to our study case,  we have four coordination settings ($k_1, k_2, k_3, k_4$).
For different settings, the observable $\bar{L}_w$ ($w \in \{x,y,z\}$) of a experimental trajectory should be different.

\subsection{Construct the testable points table}
\label{sec:testTables}
%This is sufficient
%condition for the existence of cyclic behaviors.
% {\color{red} CITE!!! that the rotational motion is of the ¡°tumbling¡± kind }
% }.

%one main point of the measurement method is that
%\begin{itemize}
%  \item
%  $L$ is coordination depended.
The testable points tables are Table \ref{tab:k1-k4} and Table \ref{tab:k1-k4p}. Lets see an example for constructing the testable points tables.

If the $R$, $P$ and $S$ are at (0,0,0), (1,0,0) and (0,1,0) respectively (setting $k_2$, see Table \ref{tab:k1-k4}), in long run, the only $\bar{L}$ components deviating from 0 with the statistic
significance should be $L_z$. Because, TASP predicts that cycles exist only in the X-Y plane along R,P,S,R.... According to $L$ definition in Eq \ref{eq:AM}, observed $L$ should have only the component on the $Z$ direction upwards. This can be shown in $k_2$ setting in Fig.~\ref{fig:k1-k4} and $k$=2 condition in Table \ref{tab:k1-k4}.
In same way, for each of the ($k_1, k_2, k_3, k_4$) setting, expected observation of $L$ can be represented as red arrows in Fig.~\ref{fig:k1-k4}. At the same time,
 % which can be test by data.
%      \item
in matrix form, testable expectations on these four settings are shown in Table \ref{tab:k1-k4} too. %and as red arrows in Fig~\ref{fig:k1-k4}.

      In summary, for the four games and four coordination setting, RPS-CH falls into 48 (3 $L$-components $\times$ 4 coordination settings $\times$ 4 games) testable points listed in the right-most three columns in Table~\ref{tab:k1-k4p}.\footnote{
          %{\color{red}
          %Take the symmetry of the matrix PRSD$_U$ and  PRSD$_S$  into account, the 48 test points can be compressed.
          Actually, disregarding the 4D$\rightarrow$3D projection, the game is 4D, cross production of two 4D vectors is an antisymmetric tensor having 6 components. Each of the 6 components is an  observable and independent. So, in four  games, only 24 test points are independent.
          %}
          %For less abstract,
          For brevity,
          the measurements and the results are presented without this compression. Regular 3-simplex (normal tetrahedron structure) representation is also suitable for a RPSD strategy game in general.
      %    But to decompose
          But decomposing
          vector $L$ in normal tetrahedron structure could lead to additional
  %        complex for visible.
          complexity to visualize.
          }
          Respectively, present the evolutionary trajectory in four coordination settings, then we measure the $L$ using Eq.~\ref{eq:AM} for the four games in in the experiments. Then, we can compare the actual motion with the three hypothesis above.

%  \item
%%      Then, by changing the reference point over full 455 points of the lattice, the robustness can be evaluated.
%\end{itemize}

%ΪÁ˲âÁ¿ËûÃǵÄÊý¾ÝÖУ¬ÊÇ·ñ´æÔÚÖÜÆÚÐԵĽṹ£¬ ÎÒÃÇ¿ÉÒÔͨ¹ýÖ±½Ó²âÁ¿ËüµÄ½Ç¶¯Á¿(Angular Momentum). ÎÒÃÇ¿ÉÒÔ½«ÖÐÐÄÉèÖÃÔÚNash ¾ùºâµãÉÏ£¬È»ºó£¬Ö±½Ó°´ÕսǶ¯Á¿µÄ¶¨Ò壬¹Û²ìÕâ¸ö²âÁ¿Á¿¡£\\
%½Ç¶ÈÁ¿µÄ´æÔÚÊÇÓÐÖÜÆÚÔ˶¯µÄ³ä·ÖÌõ¼þ\\

\begin{table}
\begin{center}
\caption{\label{tab:k1-k4} Testable TASP hypothesis on  $\bar{L}_x,\bar{L}_y,\bar{L}_z$}
\small
\begin{tabular}{|c|c|c|c|c|}
\hline
	Setting ($k$)  &~~~ $x$-$y$-$z$; $O$ ~~~& ~~~$\bar{L}_x$~~~ & ~~~$\bar{L}_y$~~~ & ~~~$\bar{L}_z$~~~ \\
\hline
	 1  & $R$-$P$-$S$; $D$ & \textbf{+} & \textbf{+} & \textbf{+} \\
\hline
	 2  & $P$-$S$-$D$; $R$ & $\bigcirc$  & $\bigcirc$ & \textbf{+} \\
\hline
	 3   & $S$-$D$-$R$; $P$ & $\bigcirc$  & \textbf{--}  & $\bigcirc$\\
\hline
	 4  & $D$-$R$-$P$; $S$ & \textbf{+} & $\bigcirc$ & $\bigcirc$\\
\hline
\end{tabular}\\
\scriptsize
Setting three of the four pure strategies along column $x$-$y$-$z$ [$e_x=(1,0,0)$, $e_y=(0,0,1)$, $e_z=(0,0,1)$], meanwhile, $O$ state assigned at (0,0,0). Testable hypotheses (PRS-CH) are in last 3 columns in which '\textbf{+}' ('\textbf{--}') or $\bigcirc$ means the $L_w$ should along (oppose to) $w$-axis direction or not deviating from 0.  \\
\end{center}
%\end{table}
%
\begin{center}
\caption{\label{tab:k1-k4p}Experimental ($\bar{L}_x,\bar{L}_y,\bar{L}_z$)$\times 10^{-3}$ in four setting }
%\scriptsize
\small
\begin{tabular}{|cc|rrr|ccc|}
\hline
	 $k$ & game & $\bar{L}_x$ & $\bar{L}_y$ & $\bar{L}_z$ &$p_x$& $p_y$ & $p_z$  \\
\hline
	 1 & 0 & 4.5 & 4.0 & 5.2 & +$^{***}$ & +$^{***}$ & +$^{***}$ \\
	 1 & 1 & 6.7 & 8.0 & 6.4 & +$^{***}$ & +$^{***}$ & +$^{***}$ \\
	 1 & 2 & 3.9 & 4.2 & 3.3 & +$^{***}$ & +$^{***}$ & +$^{**}$ \\
	 1 & 3 & 5.2 & 5.5 & 4.2 & +$^{***}$ & +$^{***}$ & +$^{***}$ \\
\hline
	 2 & 0 & 0.5 & -0.7 & 4.5 & $\bigcirc$ & $\bigcirc$ & +$^{***}$ \\
	 2 & 1 & -1.3 & 0.3 & 6.7 & $\bigcirc$ & $\bigcirc$ & +$^{***}$ \\
	 2 & 2 & -0.3 & 0.7 & 3.9 & $\bigcirc$ & $\bigcirc$ & +$^{***}$ \\
	 2 & 3 & -0.3 & 1.0 & 5.2 & $\bigcirc$ & $\bigcirc$ & +$^{***}$ \\
\hline
	 3 & 0 & 1.2 & -4.0 & 0.5 & $\bigcirc$ & --$^{***}$ & $\bigcirc$ \\
	 3 & 1 & -1.5 & -8.0 & -1.3 & $\bigcirc$ & --$^{***}$ & $\bigcirc$ \\
	 3 & 2 & -1.0 & -4.2 & -0.3 & $\bigcirc$ & --$^{***}$ & $\bigcirc$ \\
	 3 & 3 & -1.3 & -5.5 & -0.3 & $\bigcirc$ & --$^{***}$ & $\bigcirc$ \\
\hline
	 4 & 0 & 5.2 & 0.7 & 1.2 & +$^{***}$ & $\bigcirc$ & $\bigcirc$ \\
	 4 & 1 & 6.4 & -0.3 & -1.5 & +$^{***}$ & $\bigcirc$ & $\bigcirc$ \\
	 4 & 2 & 3.3 & -0.7 & -1.0 & +$^{**}$ & $\bigcirc$ & $\bigcirc$ \\
	 4 & 3 & 4.2 & -1.0 & -1.3 & +$^{***}$ & $\bigcirc$ & $\bigcirc$ \\
\hline
\end{tabular}\\
\scriptsize
Superscript [$^{(.)}$, $^{(*)}$, $^{(**)}$, $^{(***)}$]  represents $p$ less than [0.2, 0.1, 0.05, 0.01]. $p_w$-values coming from one-sample \emph{ttest} with null hypothesis that the population mean $\bar{L}_w$ is equal to 0.
The sample sizes for each test point are the total number of PPT and are [237, 217, 237, 237] for game-[0,1,2,3], respectively.
\end{center}

\begin{center}
\caption{\label{tab:LxyzZval} $|L|$ and cross game comparison for $(L_x,L_y,L_z)$ $\times 10^{-3}$ }
%\scriptsize
\small
\begin{tabular}{|c|r|c|c|c|}
  \hline
game &	$|L|$ &	   	   1 	  &	   	   2 	  &	   	   3  	  \\
% &	 $z_x,z_y,z_z$&	 $z_x,z_y,z_z$&	 $z_x,z_y,z_z$\\
  \hline
  0&  7.9 &--$^{\circ},$--$^{**},$--$^{*}$  &	 --,+,--  &	--,--,+~~    \\
 1&  12.2	&   &+$^{\circ}$,+$^{**}$,+$^{**}$&	+,+,+$^{*}$ \\
 2&  6.6	&  	&	 	&	  --,--,--~~  \\
 3&	  8.7&  	&	 	&	   \\
   \hline
\end{tabular}\\
\scriptsize
Compare (Wilcoxon rank sum) cycle strength of the game in row to the game in column. For example, the symbol (--$^{\circ}$) in 3rd column means $L_x$ in game-0  is smaller than  game-1 (at $p<0.2$). Definition of the superscript is as Table \ref{tab:k1-k4p}.
% {\color{red} need +- not 0! go back to table 1}
\end{center}
%\end{table}
%
%\begin{table}
   \begin{center}
\caption{\label{tab:k1L39L41} Time dependence of $\bar{L}^{(1,1,1)}$ $\times 10^{-3}$  }
 \small
\begin{tabular}{|c|ccrc|}
\hline
  game & $\bar{L}_{1,40}^{(1,1,1)} $ & $\bar{L}_{41,80}^{(1,1,1)} $ & $\Delta \bar{L}^{(1,1,1)} $    & Samples \\
\hline
  0  &  7.5 &  1.7 & -5.8$^{***}$     &(351, 351) \\
  1  &  7.1 &  6.9 & -0.2~~~~          &(351, 321) \\
  2 &   4.9 &  2.7 & -2.2~~~~          &(351, 351) \\
  3  &  5.1 &  4.3 & -0.8~~~~           &(351, 351) \\
\hline
\end{tabular}\\
\scriptsize
%Using $k_1$ setting, each $L_w$ [$w$=$(x,y,z)$] of a PPT projects in (1,1,1) vector forms one sample.
$(a,b)$ in Samples column indicates the samples from (1st,2nd)-half periods in the game sessions. Statistic uses \emph{ttest} with $\Delta\bar{L}$=0
 \end{center}
 \end{table}

\section{Results}\label{sec:resultsoncycles}
\subsection{Cycle existence and direction }
\label{sec:cycleDir}
\textbf{Result}: Cycles \emph{exist  and only exist}
paralleling RPS-plane in all of the 4 game experiments.
Cyclic evolutions   are along $RPSR$... in all of the 4 game experiments. \\
\textbf{Support material:}
%  The existence of cycles can be confirmed if mean $\bar{L}$ components deviate from 0 with statistic significantly. This is sufficient condition for the existence of cyclic behaviors.  In E-space,
 Statistics results of ($L_x, L_y, L_z$), from 4 settings and
 games respectively, are shown in Table~\ref{tab:k1-k4p}.
 In $k_1$ setting, the full D strategy is settle at (0,0,0).
 All the three components ($L_x, L_y, L_z$) $>$ 0 significant ($p<0.05$) for all of the 4 games.
%  because the 12 lower bound of the 95\% confident interval (for all of the 4 game=0,1,2,3  in all this three component $L_x, L_y, L_z$) are larger than 0.
  In $k_2$ setting, only $L_z > 0$ is statistically
  significant $(p<0.05)$. A straightforward interpretation
  is that cycle \emph{exists and only exist} in RPS-plane too.
  This result is supported by $k_3$ setting and $k_4$ setting.
  Comparing the theoretical expectations (Table \ref{tab:k1-k4})
  and empirical results (\ref{tab:k1-k4}),  RPS-CH is supported at all of the 48 testable points.

 %On existed cycles, its direction
 The direction of existed cycles
 can be distinguished by
 % take
 taking the signal of ($L_x,L_y,L_z$) into account.
 Empirical signal (+ or -- in  Table~\ref{tab:k1-k4p}) of ($L_x, L_y, L_z$), comparing with RPS-CH signal (+ or -- in Table~\ref{tab:k1-k4}) by  the $k$-settings and games respectively, one can find that  RPS-CH is supported excellently at all of the 48 testable points too.

\subsection{Cycle strength}
\label{sec:cycleStrength}
\textbf{Result}: Strength of cyclic motion in game-1 (unstable and High-pay) is the largest. Strength of cycles is negatively dependent on $P_D$ (average play of Dumb). \\
\textbf{Support material:}  The rotation strength of cycles can be quantified by the vector mode
  $|\bar{L}|$ $\!\equiv $ $\!(\bar{L}_x^2 + \bar{L}_y^2 + \bar{L}_z^2)^{1/2}$.
 The game-1   has the largest rotation strength $|L|$ shown in the 2nd-column of Table~\ref{tab:LxyzZval}.
This result is also supported by the  statistical test (Wilcoxon rank sum) by pair games comparison.  In Table~\ref{tab:LxyzZval}, over the 4 game, the strength orders can be compared with the arrows in Table~\ref{tab:4game}. All the results meet RPS-CH \cite{friedman2010tasp} well.\\

 Strength of cycles is negatively dependent
on $P_D$ (average play of Dumb).
At the same time,
the result in Fig.~\ref{Dumb_LxLyLzByGame} has to be explained
------ Strength of cycles is negatively dependent on $P_D$ (average play of Dumb).
This finding is statistically significant.\footnote{In session level, there is 12 samples ($n$=12) for each $L_w$. OLE test results is that the negative dependence significant with $(p_x,p_y,p_z) <(0.05,0.05,0.05)$.
This relationship can be partly interpreted by a
discrete time Logit dynamics model \cite{XZW20130621}.
In tens of dynamics (or learning) models (e.g., \cite{HofbauerSigmund2003,Camerer2003}), which could
meet all these empirical observations better is unaware.}

\subsection{Cycle persistence}\label{sec:CyclePerisistenceTASP}

\textbf{Result}: Persistence of cycle in game-1 performs best. Except game-0, persistence of cycle can not be rejected by data.\\
\textbf{Support material:}
   One way to test the persistence of cycles is to compare $L$ samples in early and latter periods.
   In session level, the hypothesis ($L_{1,40}=L_{41,80}$) can not be rejected in general.\footnote{In total 36 samples (4  games $\times$ 4 sessions/game $\times$ 3 components of $\bar{L}$),
   only one sample can be rejected ($\bar{L}_y$ in the second session in game-0, $p$=0.036$<$0.05) , or in other words the persistence cycles can not be rejected in other 35 samples.}

    At game level to test the persistence,  we can project $(L_x,L_y,L_z)$ into
    (1,1,1)-vector in $k_1$ setting to build a combinative scale $L^{(1,1,1)}$.
    For $L^{(1,1,1)}_{1,40}$ and $L^{(1,1,1)}_{41,80}$ comparison,
    both have 351 samples\footnote{The 351 samples include the samples from 3 $L$-component $\times$ 39 period/session $\times$ 3 sessions/game.
    But game 1 is 30 samples less because there are only 70 periods in its 3 sessions.
    See \cite{friedman2010tasp} for the details.}, and results are shown in Table~\ref{tab:k1L39L41}. One unexpected\footnote{Referring to TASP experiment designer's expectation, in the unstable treatments beliefs (on actions) should (more) continue to cycle.} result is: in two low-pay treatment, $L^{(1,1,1)}$ in game-0 declines significantly and more strongly than that in game-2. Nevertheless, in the four treatments,
    cycle persistence in game-1 has the best performance, which meets TASP theory again.

\section{Discussion}
\label{sec:Discuss}
%Theory is, in other words, never more than a hypothesis. When the observations of facts do not agree with a theory,
%% i.e. when they do not make sense in the frame of the theory utilised in carrying out the research,
% the theory has to be discarded and replaced by another one which promises a better fit.

In the four exemplified experiments \cite{friedman2010tasp}, firstly in this paper, the Dekel-Scotchmer cycles are reported.
We looked and
we see behavior with many of the properties the theorists \cite{benaim2009learning,dekel1992evolution,Weibull1997,Sandholm2011} told us that we would see. To the best of our knowledge, no only in Dekel-Scotchmer game, no cycle has been reported in any four strategy games before. 
These observed cycles, together with the cycles obtained in recent experiments \cite{XZW20130621,XuWang2011ICCS,Friedman2012}, we wish, could provide a novel way to merger the
%the social dynamical behaviors.
%The observed cycles,
%,XuWang2010XM,
%wang2012evolutionary,Xu2013,XuWangAsymmetryRPS,
%XWMm2011,WangXu2011,xu2012test,xu2012periodic} in their own and in others' experiments (e.g. in \cite{Binmore2001,selten2008}),
%suggest that
expected and the actual  motions.
%not only to animals with genetically heritable strategy, but also
%to human strategy interactions in laboratory.

 In experiments, cycle, as the typical non-equilibrium phenomena, have been long sought but the necessary condition for its existence is unclear \cite{Nowak2012,Friedman2012,XZW20130621}. In this view, current paper could server as the third exemplified evidence between the two condition (the continuous time and full information environments \cite{Friedman2012} and the discrete time and only local information environments \cite{XZW20130621}). In the experiment investigated here, the time is discrete but the information is full.
% That means neither information providing nor time continuation  is the only necessary condition.
  Nevertheless, the necessary conditions for cycle existence is still an open question.

%I wish this reanalysis could add knowledge to the cycles experiments \cite{Friedman2012,Nowak2012,2013arXiv1301.3238X}.

%The RPSD game was firstly suggested by Dekel and Scotchmer in 1992.  Before the testing TASP experiments in 2010~\cite{friedman2010tasp},
In closed related literatures, as mentioned \cite{Friedman2012,Nowak2012}, experimental work is surprisingly sparse. But one result, which is straightly contrasty to our results reported here, has to be reminded.
Before~\cite{friedman2010tasp}, a series of RPSD games had been tested experimentally in 1999 by Von Huyck $et.al.$ \cite{Huyck1999}.
% in which the RPS cycle was called as Dekel and Scotchmer cycles~\cite{dekel1992evolution}.
 One remarkable result is that: \emph{The subjects don't exhibit the kind of correlated behavior predicted by the dynamic} (p139 in \cite{Huyck1999}). Then, a conclusion was emphasized:  \emph{A lesson from the experiment is that one should discount models that predict deterministic cycles} (p148 in \cite{Huyck1999}).
On the contrary, referring to the cycles observed from~\cite{Friedman2012} data,
we suggest that their results on the cycle in their data \cite{Huyck1999} are worth of being rechecked.\footnote{Using the angular momentum measurement in Eq~\ref{eq:AM}, and using the projections of each PPT in (1,1,1)  as described in section \ref{sec:CyclePerisistenceTASP},
a pilot result is,
in $k_1$ setting, the RPS cycles do exist ($p<0.05$).
%This result, together with the information and payoff treatment
%effects on cycles,  need to be confirmed in more details further.
}
    %This should be helpful to answer  the open question in experimental game
    %------ what is the necessary condition for the existence of cycles \cite{Friedman2012}.
Using cycle in Dekel-Scotchmer game as the exemplified calibration, we suggest,
%the results from Cason $et.al.$ \cite{Friedman2012,friedman2010tasp} and Von Huyck $et.al.$ \cite{Huyck1999}  need to be reinvestigated. At the same time,
there has many cycles has been existed in existed experiment data.

For further investigations on the cycles of social motion, between laboratory experiments and evolutionary game theory, one central question is still: Whether the actual motions coincide with the expected motions and vise versa? As illustrated in \cite{XZW20130621,Friedman2012} the evolutionary trajectory can be calculated analytically or numerically based on a evolutionary model, so cyclic behaviors can be predicted theoretically. Together with stationary observations of a game (e.g., mean observations \cite{selten2008} and distribution in strategy space \cite{XuWang2011ICCS,Nowak2012}), observations of cycles (e.g. $L$ in this paper, frequency of cycle \cite{XZW20130621} and CRI \cite{Friedman2012}) can server as a set of calibrations to constraint game models.

\section*{Acknowledgements}
Grant of experimental social science (985-project) for Zhejiang University and SKLTP of ITP-CAS (No.  Y3KF261CJ1) support this research.

% \bibliography{RPSreference9}

\begin{thebibliography}{10}
\expandafter\ifx\csname url\endcsname\relax
  \def\url#1{\texttt{#1}}\fi
\expandafter\ifx\csname urlprefix\endcsname\relax\def\urlprefix{URL }\fi
\expandafter\ifx\csname href\endcsname\relax
  \def\href#1#2{#2} \def\path#1{#1}\fi

\bibitem{dekel1992evolution}
E.~Dekel, S.~Scotchmer, On the evolution of optimizing behavior, Journal of
  Economic Theory 57~(2) (1992) 392--406.

\bibitem{friedman2010tasp}
T.~N. Cason, D.~Friedman, E.~Hopkins, Testing the tasp: An experimental
  investigation of learning in games with unstable equilibria, Journal of
  Economic Theory 145~(6) (2010) 2309--2331.

\bibitem{nash1950equilibrium}
J.~F. Nash, Equilibrium points in n-person games, Proceedings of the national
  academy of sciences 36~(1) (1950) 48--49.

\bibitem{benaim2009learning}
M.~Bena{\=\i}m, J.~Hofbauer, E.~Hopkins, {Learning in games with unstable
  equilibria}, Journal of Economic Theory 144~(4) (2009) 1694--1709.

\bibitem{Friedman2012}
T.~N. Cason, D.~Friedman, E.~Hopkins, Cycles and instability in a
  rock-paper-scissors population game: a continuous time experiment, Review of
  Economic Studies 81 (2014) Forthcoming.

\bibitem{Huyck1999}
J.~Van~Huyck, F.~Rankin, R.~Battalio, What does it take to eliminate the use of
  a strategy strictly dominated by a mixture?, Experimental economics 2~(2)
  (1999) 129--150.

\bibitem{Samuelson2002}
L.~Samuelson, Evolution and game theory, The Journal of Economic Perspectives
  16~(2) (2002) 47--66.

\bibitem{Weibull1997}
J.~Weibull, Evolutionary game theory, The MIT Press, 1997.

\bibitem{Sandholm2011}
W.~Sandholm, Population games and evolutionary dynamics, The MIT Press, 2011.

\bibitem{hauert2002volunteering}
C.~Hauert, S.~De~Monte, J.~Hofbauer, K.~Sigmund, Volunteering as red queen
  mechanism for cooperation in public goods games, Science 296~(5570) (2002)
  1129--1132.

\bibitem{Sin96}
B.~Sinervo, C.~Lively, The rock-paper-scissors game and the evolution of
  alternative male strategies, Nature 380~(6571) (1996) 240--243.

\bibitem{Kir04}
B.~Kirkup, M.~Riley, Antibiotic-mediated antagonism leads to a bacterial game
  of rock-paper-scissors in vivo, Nature 428~(6981) (2004) 412--414.

\bibitem{Plott2008}
C.~Plott, V.~Smith, Handbook of experimental economics results, North-Holland,
  2008.

\bibitem{Friedman1996}
D.~Friedman, Equilibrium in evolutionary games: Some experimental results, The
  Economic Journal 106~(434) (1996) 1--25.

\bibitem{Huyck2008}
J.~Van~Huyck, Emergent conventions in evolutionary games, Handbook of
  Experimental Economics Results 1 (2008) 520--530.

\bibitem{Nowak2012}
M.~Hoffman, S.~Suetens, M.~Nowak, U.~Gneezy, An experimental test of nash
  equilibrium versus evolutionary stability (2012)., Nature Comm., submit.

\bibitem{berninghaus2003power}
S.~K. Berninghaus, K.-M. Ehrhart, The power of ess: An experimental study,
  Journal of Evolutionary Economics 13~(2) (2003) 161--181.

\bibitem{XZW20130621}
B.~Xu, H.-J. Zhou, Z.~Wang, Cycle frequency in standard rock¨cpaper¨cscissors
  games: Evidence from experimental economics, Physica A: Statistical Mechanics
  and its Applications~(0).

\bibitem{XuWang2011ICCS}
B.~Xu, Z.~Wang, Evolutionary Dynamical Pattern of "Coyness and Philandering":
  Evidence from Experimental Economics, Vol. VIII, p1313-1326, NECSI Knowledge
  Press, ISBN 978-0-9656328-4-3., 2011.

\bibitem{XuWang2010XM}
B.~Xu, Z.~Wang, Bertrand-edgeworth-shapley cycle in a $2 \times 2$ game, SSRN
  eLibrary, 2010 International Workshop on Experimental Economics and Finance
  Program, Xiamen, December 15-16.

\bibitem{wang2012evolutionary}
Z.~Wang, B.~Xu, Evolutionary rotation in switching incentive zero-sum games,
  arXiv preprint arXiv:1203.2591.

\bibitem{Xu2013}
B.~Xu, Cycles of strategies and changes of distribution in laboratory public
  goods games: An experimental investigation, ESA International Conference,
  2013 ESA World Meetings in Zurich Accepted.

\bibitem{XuWangAsymmetryRPS}
B.~Xu, H.~Zhou, Z.~Wang, Asymmetry spectrum of cycle amplitude in
  rock-paper-scissor game of experimental economics, ESA International
  conference 2013, http://dx.doi.org/10.2139/ssrn.2085459.

\bibitem{XWMm2011}
B.~Xu, Z.~Wang, Social transition spectrum in constant sum 2x2 games with human
  subjects, SSRN eLibrary, Social Transition Spectrum in Constant Sum 2x2 Games
  with Human Subjects, http://dx.doi.org/10.2139/ssrn.1910045.

\bibitem{WangXu2011}
Z.~Wang, B.~Xu, Spontaneous time symmetry breaking in system with mixed
  strategy nash equilibrium: Evidences in experimental economics data, Bulletin
  of the American Physical Society 56 (2011).

\bibitem{xu2012test}
B.~Xu, Z.~Wang, Test maxent in social strategy transitions with experimental
  two-person constant sum $2 \times 2$ games, Results in Physics 2~(0) (2012)
  127--134.

\bibitem{xu2012periodic}
B.~Xu, S.~Wang, Z.~Wang, Periodic frequencies of the cycles in 2$\times$2
  games, arXiv preprint arXiv:1208.6469.

\bibitem{hestenes1999new}
D.~Hestenes, New foundations for classical mechanics, Springer, 1999.

\bibitem{young2008stochastic}
H.~Young, {Stochastic Adaptive Dynamics}, New Palgrave Dictionary of Economics,
  revised edition, L. Blume and S. Durlauf, eds. Zanella, G.(2007),¡°Discrete
  Choice with Social Interactions and Endogenous Membership,¡± Journal of the
  European Economic Association 5 (2008) 122--153.

\bibitem{deng2004stationary}
M.~L. Deng, W.~Q. Zhu, Stationary motion of active brownian particles, Physical
  Review E 69~(4) (2004) 046105.

\bibitem{zhu2005stationary}
W.~Q. Zhu, M.~L. Deng, Stationary swarming motion of active brownian particles
  in parabolic external potential, Physica A: Statistical Mechanics and its
  Applications 354 (2005) 127--142.

\bibitem{deng2009some}
M.~Deng, W.~Zhu, Some applications of stochastic averaging method for quasi
  hamiltonian systems in physics, Science in China Series G: Physics, Mechanics
  and Astronomy 52~(8) (2009) 1213--1222.

\bibitem{cai2007stochastic}
G.~Cai, Y.~Lin, Stochastic analysis of time-delayed ecosystems, Physical Review
  E 76~(4) (2007) 041913.

\bibitem{HofbauerSigmund2003}
J.~Hofbauer, K.~Sigmund, Evolutionary game dynamics, Bulletin of the American
  Mathematical Society 40~(4) (2003) 479.

\bibitem{Camerer2003}
C.~Camerer, R.~S. Foundation, Behavioral game theory: Experiments in strategic
  interaction, Vol.~9, Princeton University Press Princeton, NJ, 2003.

\bibitem{selten2008}
R.~Selten, T.~Chmura, Stationary concepts for experimental 2x2-games, The
  American Economic Review 98 (2008) 938--966.

\end{thebibliography}
 \end{document}